\documentclass[aps,prl,reprint,amssymb,superscriptaddress]{revtex4-1}
\usepackage{graphicx}
\usepackage{amsmath}
\usepackage{mathtools}

\usepackage{color}
\usepackage{enumitem}
\usepackage{graphicx}
\usepackage[normalem]{ulem}
\usepackage{comment}

\newif\ifHighlitedChanges
\def\ifHighlitedChanges{\iftrue}
\ifHighlitedChanges
  
  \def\STRIKE#1{{\color{red}\sout{#1}}}
\else
  
  \def\STRIKE#1{\relax}
\fi

\begin{document}

\bibliographystyle{apsrev}

\title{Electron-Transfer-Induced Thermal and Thermoelectric Rectification}
\author{Galen T. Craven}  
\affiliation{Department of Chemistry, University of Pennsylvania, Philadelphia, PA  19104, USA} 
\author{Dahai He}
\affiliation{Department of Physics and Jiujiang Research Institute, Xiamen University, Xiamen 361005, China} 
\author{Abraham Nitzan}
\affiliation{Department of Chemistry, University of Pennsylvania, Philadelphia, PA  19104, USA} 
\affiliation{School of Chemistry, Tel Aviv University, Tel Aviv 69978, Israel}

\begin{abstract}
Controlling the 
direction 
and 
magnitude 
of both heat and electronic currents using
rectifiers has significant implications 
for the advancement of 
molecular circuit design.
In order to facilitate the implementation of new transport phenomena in such molecular structures, 
we examine thermal and thermoelectric rectification effects that 
are induced by an electron transfer process that occurs across a temperature gradient between molecules.
Historically, the only known heat conduction mechanism able to generate thermal rectification 
in purely molecular environments is phononic heat transport.
Here, we show that electron transfer between molecular sites with different local temperatures can also generate a thermal rectification effect 
and that electron hopping through molecular bridges connecting metal leads at different temperatures gives rise to asymmetric Seebeck effects, that is, thermoelectric rectification, in molecular junctions.
\vspace{0.22cm}
\begin{description}
\item[Cite as]
G. T. Craven, D. He, and A. Nitzan, \textit{Phys. Rev. Lett.} \textbf{121}, 247704 (2018)
\item[DOI]
 \href{dx.doi.org/10.1103/PhysRevLett.121.247704}{10.1103/PhysRevLett.121.247704}
\end{description}
\end{abstract}

 \maketitle
Rectifiers are devices that 
promote preferential current flow in one direction.
The development and application of single-molecule and solid-state electronic rectifiers 
has been integral in the advancement of both nanoscale and macroscale electrical circuit designs \cite{Ratner1974rectifier,Streetman2005solidstate,Bredas2007,Xiang2016,Plett2017}.
Thermal rectifiers,
which  
facilitate 
unidirectional heat flow,
have also been developed and these devices
have potential applications in diverse technologies including 
thermal management systems for nanoscale electronics,
energy-harvesting devices, 
and thermal circuits 
which use heat instead of electricity to perform operations \cite{Li2012,Sato2012, Maldovan2013}.
In molecular environments where heat conduction is dominated by vibrational (phononic) energy transfer, thermal rectification may result from nonlinear coupling, that is, anharmonicity, and the breaking of symmetry.
Such phenomena have been the focus of significant theoretical and experimental examination \cite{Terraneo2002,Li2004,Segal2005prl,Chang2006,Segal2008prl,Strano2015,Wang2017thermaldiode}.
Phononic heat conduction has been discussed as a possible mechanism for the operation of thermal components such as
thermal transistors \cite{Li2006,Ben-Abdallah2014,Joulain2016},  thermal memory \cite{Li2008}, and thermal logic gates \cite{Li2007logic}.
Thermoelectric rectification effects have also been predicted in solid-state devices
where they are
applied to induce and control electrical currents using thermal sources \cite{Kuo2010,Zhang2012}.

Recently, a new mechanism for heat transfer was identified in molecular systems undergoing electron transfer (ET) events \cite{craven16c}. It has been shown that electron hopping between molecular sites of different local vibrational temperatures is accompanied by heat transfer even when the net electronic current vanishes. Such electron-transfer-induced heat transport (ETIHT) can make a substantial contribution to the heat transfer between sites that are characterized by large reorganization energies (i.e., strong electron-phonon couplings) \cite{craven17e} and was also shown to lead to a thermal transistor effect in a properly engineered donor-acceptor site geometry \cite{craven17b}.

In this Letter, 
we show that 
the transfer of electrons between molecular donor-acceptor sites
of different temperatures
can induce thermal rectification
when the two sites are characterized by different reorganization energies,
and that the electronic conduction intrinsic to this process results in thermoelectric rectification in molecular nanostructures.
The underlying ET process between charge transfer sites is described by a Marcus formalism \cite{Marcus1956,Marcus1993} (see also Levich \cite{Kuznetsov1999Electron} and Hush \cite{Hush1958,Hush1961}) that is augmented to treat the case of ET across a thermal gradient \cite{craven16c}. 
This theory is based on a linear response (harmonic) model of the thermal environment. 
Such a harmonic environment cannot, by itself, cause rectification of heat transport. 
Here, the coupling of vibrational modes to the electronic subsystem is shown to provide the nonlinear component needed for rectification behavior to be present.
Thus, there are two major findings in this Letter: 
(a) observation that 
electron transfer events across thermal gradients in purely molecular environments
can generate significant thermal rectification
and 
(b) demonstration that such systems can show thermoelectric rectification in molecular junctions operating under zero current conditions.


\begin{figure}[]
\includegraphics[width = 8.6cm,clip]{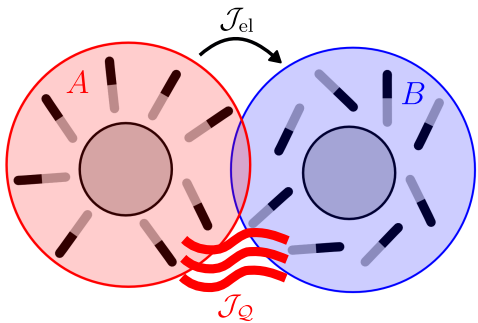}
\caption{\label{fig:Graphic_Abstract}
Schematic diagram for electron transfer between molecular charge transfer sites (gray circles) which have local solvent environments
 $A$ and $B$ (transparent ovals). The environments have respective temperatures $T_A$ (red) and $T_B$ (blue). 
The electron flux $\mathcal{J}_\text{el}$ between sites generates a heat current $\mathcal{J}_\mathcal{Q}$ between environments.
}
\end{figure}

The general formalism we apply follows the results of Ref.~\citenum{craven16c},
adapted to show the rectification effects examined here.
The first system we consider comprises two molecular sites which have respective local solvent environments $A$ and $B$ (see Fig.~\ref{fig:Graphic_Abstract}). 
Electron transfer occurs between these two sites via a hopping mechanism. 
The system has two electronic states: $a$ and $b$, which respectively correspond to excess electron localization 
on the molecular charge transfer site in environments $A$ and $B$, i.e., $a \equiv \{A^-, B\}$
and $b \equiv \{A, B^-\}$. 
Each environment contains a set of nuclear vibrational modes that are affected by electron localization 
on the charge transfer site in that environment. 
Note that, herein, we use $A$ and $B$ to denote the respective environment and the set of modes that belong to that environment.
The polaron response of the nuclear vibrational modes to the electron localization on each of the 
sites is described using the Marcus formalism for ET in which the energy surface for electronic state $m \in \left\{a,b\right\}$ is
\begin{equation}
\label{eq:energysurf}
E_m\left(x_1,x_2,\ldots,x_N\right) = \frac{1}{2} \sum_{j \in A \cup B}\!\!\kappa_j\left(x_j - \lambda^{(m)}_j\right)^2 + E'_m, \\[1ex]
\end{equation}
where $x_j$ and $\kappa_j$ are, respectively, the coordinate and
force constant of the $j$th mode,
$E'_m$ is the electronic origin of state $m$, 
and $\lambda^{(m)}_j$ is a polaron-induced shift from equilibrium of mode $j$ due to the electron localization.
The reorganization energy of each mode, 
which is a measure of the magnitude of the electron-phonon interaction, 
is $E_{\text{R}j} = \tfrac{1}{2} \kappa_j (\lambda^{(a)}_j-\lambda^{(b)}_j)^2$ \cite{note2,Nitzan2007jpcm}.
Finally, $\Delta E_{ab} = -\Delta E_{ba} = E'_a-E'_b$
is the difference in the energy origins of the states,
that is, the reaction free energy.

Our model considers ET between two sites embedded in environments with different local temperatures
$T_A = T-\Delta T/2$ and $T_B = T+\Delta T/2$
(with $\Delta T = T_B -T_A$).
The phonons that respond strongly to the difference between the charge distributions 
in the two electronic states are localized near sites $A$ and $B$ and are assumed to be equilibrated at the temperature of the local environment \cite{note3}.
To examine thermal rectification effects in this system we consider
two thermal bias states:
$\Delta T > 0$ (denoted by ``$+$'') and $\Delta T < 0$ (denoted by ``$-$'').

The ET rate between states can be derived in the strong electron-phonon coupling limit using 
a bithermal ($T_A$ and $T_B$) formulation of the Marcus semiclassical transition state theory \cite{craven16c},
or, equivalently, Fermi's golden rule evaluated in the strong coupling/high temperature limit \cite{note1,Lin1966,Lin2002Fermi,Nitzan2006chemical}. 
In this limit, nuclear tunneling is ignored and the bithermal ET rate in the corresponding thermal bias state ($\pm$) takes the form 
\begin{align}
\label{kabflux}
\nonumber k^\pm_{m \to n}  &= \frac{|V_{mn}|^2}{\hbar} \sqrt{\frac{\pi}{k_\text{B}\left(T_A  E_{\text{R}A}  +  T_B   E_{\text{R}B} \right)}} \\[0ex]
& \quad \times \exp{\left[ -\frac{\left(\Delta E_{nm} + E_\text{R}\right)^2}{4 k_\text{B} \left(T_A  E_{\text{R}A}  +  T_B  E_{\text{R}B}\right)} \right]},
\end{align}
where $V_{mn}$ is the coupling between energy surfaces, $k_\text{B}$ is Boltzmann's constant, 
$E_{\text{R}A} = \sum_{j \in A} E_{\text{R}j}$ and $E_{\text{R}B} = \sum_{j \in B} E_{\text{R}j}$ are the partial reorganization energies associated with the modes in environments $A$ and $B$, respectively,
and $E_\text{R} = E_{\text{R}A} + E_{\text{R}B}$ is the total reorganization energy for the electronic transition.

The structure of Eq.~(\ref{kabflux}) implies that the electron transfer rate is not symmetric in the two thermal bias states
for $E_{\text{R}A} \neq E_{\text{R}B}$.
The asymmetry between the partial reorganization energies of environments $A$ and $B$ can be expressed in terms of the parameter
\begin{equation}
\alpha = \frac{E_{\text{R}A}}{E_{\text{R}B}}.
\end{equation}
When $\alpha = 1$, the system is symmetric and the total reorganization energies of each environment are equal. 
When $0 \leq \alpha \neq 1$ the system is asymmetric and in this case we expect to observe
thermal rectification effects.
The asymmetrical property can arise from several sources: (a) the presence of 
more modes that participate in the ET process in one environment than the other, (b) different electronic 
characteristics of the solvent about each charge transfer site,
(c) the extent of electron localization on the two sites, 
 and/or (d) the temperature dependence of the reorganization energies \cite{Voth1999,Derr1999}.

Now, consider the energy change in each environment 
that is induced by the ET process.
To examine this heat current $\mathcal{J}_\mathcal{Q}$
we first consider the probabilities $p^\pm_a$ and $p^\pm_b$ for the system to be found in the respective electronic states $a$ or $b$
under the corresponding positive or negative thermal bias.
The time-evolution of the occupancy probabilities obey the kinetic equations
$\dot{p}^\pm_a = -\dot{p}^\pm_b =  - k^\pm_{a \to b} p^\pm_a+ k^\pm_{b \to a} p^\pm_b$.
At steady state, the net electronic current vanishes and
the unidirectional electron flux is $\mathcal{J}^\pm_\text{el} = k^\pm_{a \to b} p^\pm_a =  k^\pm_{b \to a} p^\pm_b$. 
The steady state heat current for the corresponding thermal bias state: 
\begin{equation}
\label{eq:heatfluxa}
\mathcal{J}_\mathcal{Q}^{\pm} = \mathcal{J}_\text{el}^{\pm}\frac{2(T_B-T_A)  E_{\text{R}A} E_{\text{R}B}}{T_A E_{\text{R}A}  +  T_B E_{\text{R}B}},
\end{equation}
is a product of the electron flux $\mathcal{J}_\text{el}^{\pm}$ and the energy change generated by each ET event
\cite{craven16c,craven17b,note11}.
Equation~(\ref{eq:heatfluxa}) implies that in a system with nonvanishing reorganization energies,
the heat current generated by the transfer of electrons is nonzero for $T_A \neq T_B$.
In the case of $\alpha \neq 1$, $|\mathcal{J}_\mathcal{Q}^{\pm}(\Delta T)| \neq |\mathcal{J}_\mathcal{Q}^{\pm}(-\Delta T)| \, \forall \, \Delta T \neq 0$ which implies that in asymmetric electron transfer reactions 
between molecules in environments with different local temperatures, 
thermal rectification effects can be induced solely from the
transfer of electrons.

\begin{figure}[t]
\includegraphics[width = 8.6cm,clip]{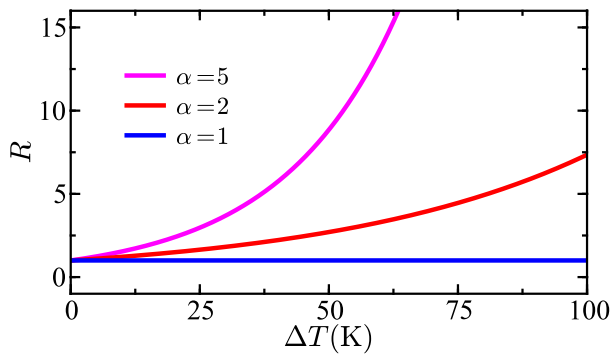}
\caption{\label{fig:Ratio_DeltaT}
Thermal rectification ratio as a function of $\Delta T$ for 
various values of $\alpha$.
Parameters are $\Delta E_{ab} = 0.5\,\text{eV}$, $E_{\text{R}A} = 0.15\,\text{eV}$, $E_{\text{R}B} = \alpha E_{\text{R}A}$, and $T = 300\,\text{K}$.
}
\end{figure}

It is clear from Eq.~(\ref{eq:heatfluxa}) that in general $\mathcal{J}^+_\mathcal{Q} \neq \mathcal{J}^-_\mathcal{Q}$. 
This asymmetry is analogous to that reported for an anharmonic vibrational mode, specifically a two-level 
system, bridging two bosonic reservoirs with different temperatures \cite{Segal2005prl}.
The magnitude of the rectification can be quantified through the thermal rectification ratio
\begin{equation}
\label{eq:ratio}
R  = \left|\frac{\mathcal{J}_\mathcal{Q}^{+}}{\mathcal{J}_\mathcal{Q}^{-}}\right|.
\end{equation}
Figure~\ref{fig:Ratio_DeltaT} illustrates rectification ratios generated by electron-transfer-induced heat transport 
between molecular charge transfer sites over variation of $\Delta T$.
For $\alpha = 1$, Eq.~(\ref{eq:ratio}) gives the expected result that 
$R = 1$.
For $\alpha = 2$, $R > 1$ for $\Delta T >0$ illustrating that thermal rectification is present 
in asymmetric systems with temperature gradients.
Note that for $\alpha \neq 1$, $R$ varies nonlinearly in $\Delta T$.
Comparing the results for $\alpha = 2$ and $\alpha = 5$, it can be observed that $R$ is also nonlinear in
$\alpha$ and, thus, for constant $\Delta T$,  small variations in the asymmetry of the system can 
result in large changes in the observed rectification ratio.
It is of interest that the observed thermal rectification effects
are generated using the harmonic surfaces given in Eq.~(\ref{eq:energysurf}). 
This is in contradiction with the traditional posit, 
arising from the theories that describe phononic thermal diodes, that thermal rectifiers must be anharmonic.

\begin{figure}[t]
\includegraphics[width = 8.6cm,clip]{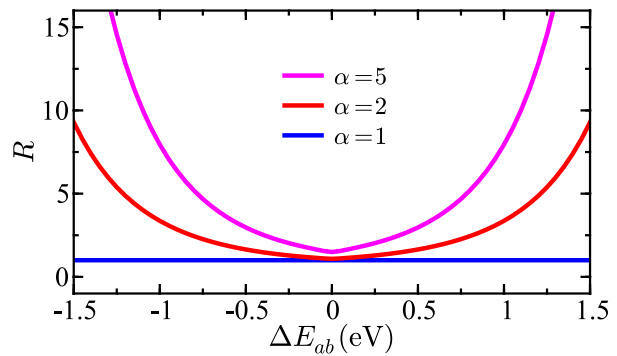}
\caption{\label{fig:Ratio_DeltaE}
Thermal rectification ratio as a function of $\Delta E_{ab}$ for 
various values of $\alpha$.
Parameters are $E_{\text{R}A} = 0.15\,\text{eV}$, $E_{\text{R}B} = \alpha E_{\text{R}A}$, $\Delta T = 25\,\text{K}$, and $T = 300\,\text{K}$.
}
\end{figure}

The thermal rectification ratios depend strongly on the reaction free energy 
$\Delta E_{ab}$ which appears implicitly in Eq.~(\ref{eq:heatfluxa}) through the $\mathcal{J}^\pm_\text{el}$ term.
Shown in Fig.~\ref{fig:Ratio_DeltaE} are rectification ratios as a
function of $\Delta E_{ab}$ with $\Delta T$ held constant.
In the $\alpha  = 2$ curve, note that $R$ is nonlinear over variation of $\Delta E_{ab}$
and also symmetric with respect to the sign of $\Delta E_{ab}$. 
This implies that the magnitude of the free energy difference $|\Delta E_{ab}|$ 
is the pertinent quantity in the rectification ratio and not its direction 
($\Delta E_{ab}$ or $-\Delta E_{ab}$) with respect to the applied temperature difference $\Delta T$.

\begin{figure}[]
\includegraphics[width = 8.6cm]{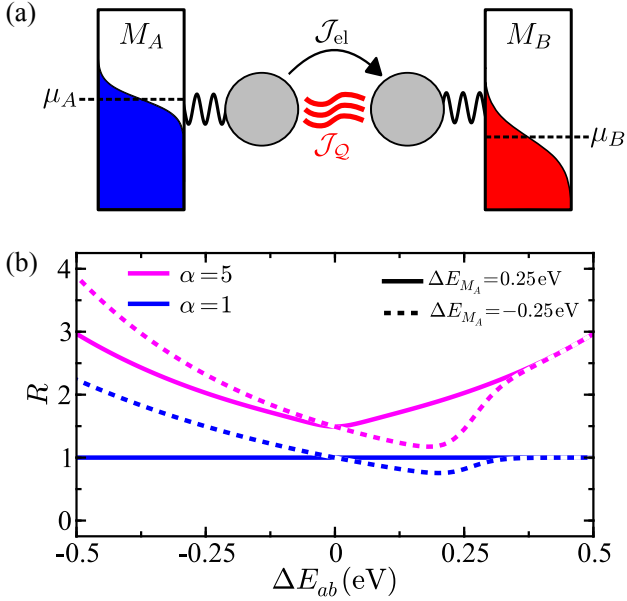}
\caption{\label{fig:Seebeck}
(a) Schematic diagram for electron and heat transport between 
molecules (gray circles) that are connected to respective metals 
$M_A$ and $M_B$ with temperatures $T_A$ (blue) and $T_B$ (red).
The chemical potential of each metal is 
shown as a dashed line. 
(b) Thermal rectification ratio as a function of $\Delta E_{ab}$ under zero electronic current conditions for various
values of $\alpha$ and $\Delta E_{M_A}$.
Parameters are $E_{\text{R}A} = 0.15\,\text{eV}$, $E_{\text{R}B} = \alpha E_{\text{R}A}$,
$\Delta T = 25\,\text{K}$, 
and $T = 300\,\text{K}$.
}
\end{figure}

While the observation just made is not easily amenable to experimental observation, 
its implication in a junction environment suggests a possible experimental demonstration as described below.
Consider the model molecular junction shown in Fig.~\ref{fig:Seebeck}(a)
in which the previous model is augmented by
placing the two molecular sites as a bridge
between two metal electrodes, $M_A$ and $M_B$,
which are
characterized by respective equilibrium temperatures $T_A = T - \Delta T / 2 $ and $T_B = T + \Delta T / 2$
and chemical potentials 
$\mu_A  = \mu + e\Phi/2$ and $\mu_B = \mu - e\Phi/2$, 
with $\Phi$ being the voltage bias across the junction \cite{note1}.
Each molecule in the bridge is 
taken to be in thermal equilibrium at the temperature of the corresponding metal to which it is coupled.
The system has three electronic states: 
$a \equiv \{A^-, B\}$,
$b \equiv \{A, B^-\}$,
and $M \equiv \{A, B\}$,
where $M$ denotes the state 
in which an excess electron is not found in the molecular bridge. 
The reaction free energies for electron insertion 
into electrodes $M_A$ and $M_B$ are $\Delta E_{M_A} = E'_M - E'_a$ 
and $\Delta E_{M_B} = E'_M - E'_b$, respectively,
where  $E'_M = \mu$ is the energy origin of state $M$
and $\Delta E_{M_A} - \Delta E_{M_B} = \Delta E_{ba}$.
The kinetic scheme for this system is \cite{Lehmann2002,Beratan2012}
\begin{equation}
M  
\xrightleftharpoons[k^\pm_{a \to M}]{k^\pm_{M \to a}}
a
\xrightleftharpoons[k^\pm_{b \to a}]{k^\pm_{a \to b}}
b 
\xrightleftharpoons[k^\pm_{M \to b}]{k^\pm_{b \to M}}
M
\end{equation}
where the interfacial molecule-metal ET rate constants are calculated 
from Marcus theory \cite{Marcus1965,Nitzan2006chemical,note1}
and the bithermal rate constants for ET between the molecular sites are taken from Eq.~(\ref{kabflux}).

The unidirectional electron flux between molecules in the forward $A \to B$ 
and reverse $B \to A$ directions for 
the corresponding thermal bias states are $\mathcal{J}_{a \to b}^{\pm}$ and $\mathcal{J}_{b \to a}^{\pm}$.
In the zero current case, 
the electron-transfer-induced heat current in the junction takes a functional form
analogous to (\ref{eq:heatfluxa}) 
where now $\mathcal{J}^\pm_\text{el} = \mathcal{J}_{a \to b}^{\pm} = \mathcal{J}_{b \to a}^{\pm}$.
The condition of the electronic current $I$ to be zero while the bias is maintained is given by the condition
$\mathcal{J}_{a \to b}^{\pm}(\Phi) - \mathcal{J}_{b \to a}^{\pm}(\Phi) =0$.
It is important to note that, unlike the Landauer case, zero electric current does not imply zero heat current.
The voltage bias 
leading to $I =0$ 
for a given temperature difference can be expressed as \cite{Tomak2004,Sanchez2013}
\begin{equation}
\label{eq:Seebeck}
\Phi = -\sum^\infty_{j = 1}  S^\pm_j \Delta T^j,
\end{equation}
where $S^\pm_k$ is the $k$th order Seebeck coefficient
in the corresponding thermal bias state, with
$S_1$ being the standard linear Seebeck coefficient.
In contrast to coherent transport, here,
$S_k^+$ can be different from $S_k^-$.
Next, we show numerical examples of 
the thermal and thermoelectric rectification phenomena associated with this model.

Electron-transfer-induced thermal rectification ratios 
in the junction with $I^\pm = 0$ are shown in the Fig.~\ref{fig:Seebeck}(b) as
a function of $\Delta E_{ab}$ for various values of $\alpha$ and $\Delta E_{M_A}$ \cite{note1}. 
The observed rectification effects arise from relations between the reorganization energies, the reaction free energies  
of both the interfacial and molecule-to-molecule ET events, and the zero-current voltage bias. 
For $\alpha = 1$, 
the rectification ratio $R$ can be different from unity,
which is observed prominently in the case $\Delta E_{M_A} = -0.25\,\text{eV}$ but is almost absent for $\Delta E_{M_A} = 0.25\,\text{eV}$ \cite{note10}.
This implies that the nonlinearity and asymmetry required to induce thermal rectification 
can arise solely from the electronic structure of the molecular bridge (expressed through $\mathcal{J}^\pm_\text{el}$), and that rectification can occur even when the reorganization energies are the same in each environment, which is not the case for the molecule-molecule system considered earlier.
Also note that when $\Delta E_{M_A} <0$ ($\Rightarrow E'_a > \mu$), $R(\Delta E_{ab}) \neq R(-\Delta E_{ab})$
for all shown values of $\alpha$,
which differs from the molecule-molecule thermal rectification ratios
shown in Fig.~\ref{fig:Ratio_DeltaE} which depend only on the magnitude of $\Delta E_{ab}$ \cite{note9}.
For $\Delta E_{M_A} > 0$ ($\Rightarrow E'_a < \mu$) which is shown using solid curves,
$R(\Delta E_{ab}) \approx R(-\Delta E_{ab})$
for both symmetric ($\alpha =1$) and asymmetric ($\alpha = 5$) environments.

\begin{figure}[t]
\includegraphics[width = 8.6cm]{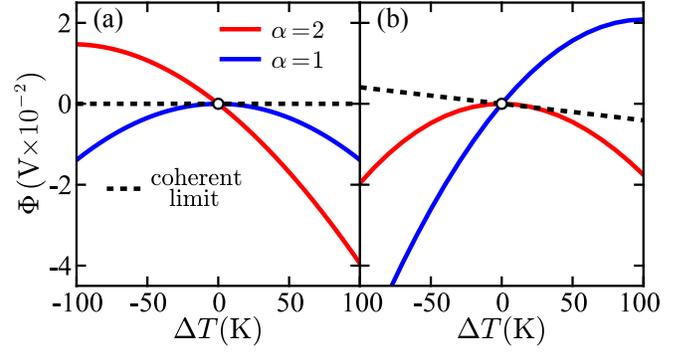}
\caption{\label{fig:Thermoelectric_Rectification}
Voltage bias $\Phi$ as function of $\Delta T$ under zero electronic current conditions for various
values of $\alpha$.
The circular marker in each panel denotes the point where $\Phi = \Delta T = 0$.
Parameters are $E_{\text{R}A} = 0.15\,\text{eV}$, $E_{\text{R}B} = \alpha E_{\text{R}A}$,
$T = 300\,\text{K}$, $V_{ab} = 0.01 \,\text{eV}$, with (a) $\Delta E_{M_A} = - \Delta E_{M_B} =  0.25 \,\text{eV}$ 
and (b) $\Delta E_{M_A} = - 1/2 \Delta E_{M_B} =  0.25 \,\text{eV}$ \cite{note1}.
}
\end{figure}

The rectification of thermoelectric response in the molecular junction model is demonstrated in Fig.~\ref{fig:Thermoelectric_Rectification}.
Shown are the values of $\Phi$ that 
lead to $I^\pm = 0$ as a function of $\Delta T$ for various values of $\alpha$ with $\Delta E_{ab}$ held constant. 
Each panel of the figure shows the results for different values of $\Delta E_{M_A}$ and $\Delta E_{M_B}$. 
It is of significance that for incoherent transport $\Phi(\Delta T) \neq -\Phi(-\Delta T) \, \forall \, \Delta T \neq 0$,
which states that when the temperature bias is reversed, the voltage bias leading to zero current does not simply change sign.
This is in contrast to the coherent limit of transport calculated using the Landauer formalism (shown by a dashed black line) where $\Phi(\Delta T) = -\Phi(-\Delta T)$
as long as the  transmission coefficient remains constant \cite{note1, Nitzan2006chemical}. 
This asymmetry with respect to $\Delta T$ implies that $S^+_k \neq S^-_k$ for even values of $k$. 
Moreover, in Fig.~\ref{fig:Thermoelectric_Rectification}(a) for 
$\alpha  = 1$ and Fig.~\ref{fig:Thermoelectric_Rectification}(b) for 
$\alpha  = 2$, $\Phi(\Delta T)  \approx \Phi(-\Delta T)$
which expresses the counter-intuitive result that when the thermal bias state is reversed, the voltage bias required to 
generate zero electronic current can remain unchanged for bridges with certain electronic properties.
It can also be observed in Fig.~\ref{fig:Thermoelectric_Rectification}(b) that
the sign of the first-order Seebeck coefficient $S_1$ for incoherent transport can be different than for coherent transport.
This implies that the direction of thermoelectric current can be controlled by altering the electronic properties of the junction environment, i.e., the reorganization energies.


In conclusion, we have shown that  thermal 
rectification effects
can be induced by the transfer of electrons 
between molecular sites that are seated in environments with 
different local temperatures. 
The electron-transfer-induced thermal rectification mechanism
differs from the standard phononic heat transport mechanism
which has historically been assumed to be the sole heat conduction channel in purely molecular systems.
We have also demonstrated that the intrinsic electronic conduction 
that accompanies this heat transfer can
give rise to thermoelectric rectification in a model molecular junction.
The treatment of electron-transfer-induced thermal and thermoelectric rectification effects in molecular junctions and at molecule-electrode interfaces will provide insight into how thermal gradients affect electron scattering and charge currents in molecular electronic devices.

Electron-transfer-induced thermal rectification effects could 
be observed experimentally by controlling the electrical currents in molecular electronic diodes \cite{Ratner1974rectifier} using thermal gradients,
and also by examining systems with suppressed phononic heat currents such as $\pi$-stacked molecular junctions \cite{Solomon2014,Li2017} and molecular donor-acceptor dyads \cite{Batista2017} with weak intramolecular vibrational couplings.
In the case of molecule-to-molecule electron transfer (see Fig.~\ref{fig:Graphic_Abstract}), 
measuring heat changes in each molecule's environment while controlling the local temperature
will require the implementation of recent advances in nanoscale thermometry \cite{Sadat2010,Menges2012, Menges2016,Mecklenburg2015}.
For molecular junctions (see Fig.~\ref{fig:Seebeck}(a)), 
the phenomena predicted in this Letter
could be measured by employing similar experimental techniques to those that have been previously
used to probe the electronic \cite{Venkataraman2015}, thermoelectric \cite{Reddy2007,Tan2011}, and thermal 
\cite{Lee2013,Kim2014,Cui2017} properties of such systems.


The research of A.N. is supported by the Israel-U.S. Binational Science Foundation, 
the German Research Foundation (DFG TH 820/11-1), 
the U.S. National Science Foundation (Grant No. CHE1665291),
and the University of Pennsylvania.
D.H. is supported by NSFC of China (Grants No. 11675133 and No. 11335006), NSF of Fujian Province (Grant No. 2016J01036).

\bibliography{j,electron-transfer,heat-transport,nonequilibrium,ratetheory,osc-bar,c7_Arxiv,craven}

\begin{thebibliography}{60}
\bibitem[{not({\natexlab{a}})}]{note2}
\bibinfo{note}{This model is equivalent to the standard polaron model
  \cite{note1}.}
\bibitem[{not({\natexlab{b}})}]{note3}
\bibinfo{note}{We furthermore assume that the distance between the sites is
  large enough such that the contribution from phononic heat transfer between
  the two sites vanishes in the harmonic approximation.}
\bibitem[{not({\natexlab{d}})}]{note11}
\bibinfo{note}{The typical magnitude of the electron-transfer-induced heat
  current in a single-molecule device is in the range $\mu \text{eV} /
  \text{ps}$ to $\text{m} \text{eV} / \text{ps}$ \cite{craven17e}. The heat
  transferred per electron is $\sim 0.01 \, \text{eV}$.}
\bibitem[{not({\natexlab{e}})}]{note10}
\bibinfo{note}{In general, the magnitude of thermal rectification will be large
  in systems where biasing of the chemical potentials significantly alters the
  molecule-to-molecule electronic flux, and will be small in the opposite
  limit.}
\bibitem[{not({\natexlab{f}})}]{note9}
\bibinfo{note}{This effect arises because constant $\Delta E_{M_A}$ implies
  constant $E'_a$ which means that variation of $\Delta E_{ab}$ is synonymous
  with variation of $E'_b$, and thus the energy gaps between the biased
  chemical potential at each electrode and the corresponding molecular energy
  level are not symmetric for $\Delta E_{ab}$ and $-\Delta E_{ab}$.}
\end{thebibliography}
\end{document}